# Dual Toffoli and Peres Reversible Gates

Claudio Moraga

*Faculty of Computer Science, Technical University of Dortmund, Germany*
*Department of Informatics, Technical University "Federico Santa María", Valparaíso, Chile*

**Abstract:** *The paper introduces dual Toffoli and Peres reversible gates, which operate under disjunctive control, and shows their functionality based on the Barenco et al. quantum model. Both uniform and mixed polarity are considered for the controls. Rewriting rules are presented, which provide a possible reduction of the number of gates and quantum cost of reversible (sub)circuits using standard Toffoli or Peres gates. Finally, a Clifford+T realization of a dual Toffoli and a dual Peres gate is shown, which may be used when mapping reversible circuits to the IBM quantum computers.*

**Keywords:** *Reversible gates; dual Toffoli gates; rewriting rules.*

## I. Introduction

One of the earliest contributions to the development of reversible/quantum circuits is due to T. Toffoli [30], [31], who proposed a functionally complete reversible circuit, that soon became known as "Toffoli gate", distinguishing two control bits and a target bit, preserving the control bits and inverting the target bit when the conjunction of the control bits became true. This reversible gate has intensively been used ever since and has received several "extensions", like multi-controlled Toffoli gates and their decomposition as V-shaped cascades of elementary Toffoli gates and ancillary bits [5], a quantum realization model [5]. mixed polarity controlled Toffoli gates [14], [26], and, in recent times, Clifford-T realizations [1] and mappings to the IBM QX quantum computers [1], [3]. Together with the Toffoli gate, the NOT gate and CNOT, the controlled NOT gate, are the basic components of reversible circuits. Their symbolic representation is shown in Fig. 1.

The realization of minimal (irreversible) binary circuits is known to be in NP [32]. Due to the constraints imposed by reversibility, like no feedback and no fan-out of gates, it is assumed that the synthesis of minimal reversible/quantum circuits is also in NP. The synthesis of reversible/quantum circuits is, therefore, mostly based on heuristics [4], [7], [8], [9], [11], [12], [15], [17], [25], [27], [28], [29]. Post-processing optimization of circuits has been applied, mainly using Templates [11], [24] and rewriting rules [26]. One of the most frequently mentioned alternative for the realization of reversible/quantum gates is based on ion traps, after the seminal work of A. Muthukrishnan and C. Stroud [22].

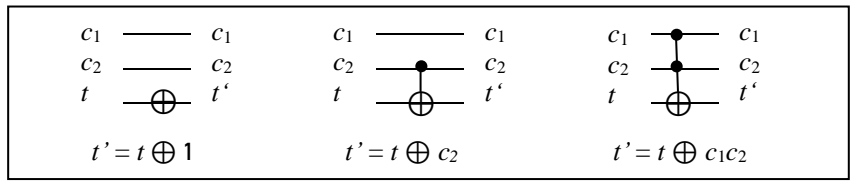

**Fig. 1**: Basic reversible gates: NOT, CNOT, and Toffoli

In the present paper Toffoli gates with disjunctive control [18], [19] will be presented, including mixed polarity. This kind of Toffoli gates will be called *dual*. Similarly for the case of Peres reversible gates. Some authors call them simply OR-Toffoli and OR-Peres gates. (*e.g.* [2], [18]). Rewriting rules will be developed for the Post-processing of circuits, allowing to replace, when applicable, sub-circuits based on classical reversible gates by simpler circuits using dual reversible gates.

## II. Formalisms

*Definition 1*: A disjunct controlled dual Toffoli gate has the following behavior: the target bit will be inverted iff the *disjunction* of the binary control signals is true, i.e. if any or both control bits have the value 1. The gate remains inhibited if *both* controls have the value 0.

Its symbol, its quantum model under positive polarity, (similar to the Barenco *et al.* [5] model) and its specification matrix [18] are shown in Fig 2, where the connection between a control bit and the target inverter is represented by a triangular symbol ▼, (in black when the expected control signal is 1), which is close to the disjunction sign ∨ of the Mathematical Logic. In the classical reversible gates the connection between a control bit and the target is represented by a black dot if the activating control signal has the value 1 or a white dot if the activating control signal has the value 0 [14], [26]. In what follows, as a matter of fairness, the quantum model will be called Barenco *et al.* model. This model shows that the dual Toffoli gate, as the classical Toffoli gate, has a quantum cost of 5.

In the Barenco *et al.* model, the matrix specification of the **V**-gate is the square root of the matrix specification of the NOT gate [5]. Therefore **V·V** = NOT.

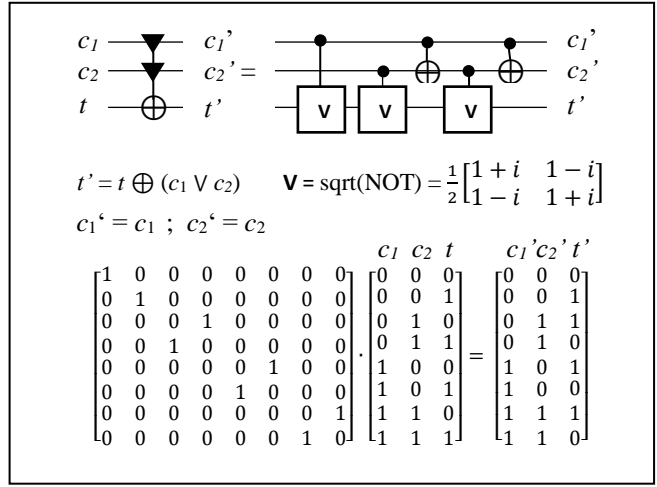

**Fig.2**: Symbol, Barenco *et al*. quantum model, and matrix specification of the basic disjunct controlled dual Toffoli gate.

It is simple to see in the Barenco *et al.* model of Fig. 2 that if $c_1$ has the value 1 ("$c_1$ is 1") and $c_2$ has the value 0 ("$c_2$ is 0") the first **V**-gate will become active, and the second one will be inhibited, thus behaving as the identity. Furthermore, $c_1$ will activate the CNOT gates, producing a "local 1" that will activate the third **V**-gate. Finally, the cascade of the two active **V**-gates produce the expected NOT behavior. (Notice that the last CNOT gate only recovers the original value of $c_2$).

In the case that $c_1$ is 0 and $c_2$ is 1, the first **V**-gate and the CNOT gates will be inhibited, whereas the second and the third **V**-gates will be activated and their product will produce the expected NOT behavior.

Finally, if both $c_1$ and $c_2$ are 1, the two first **V**-gates become activated and produce the expected NOT behavior. Since the CNOT gates will also become activated by $c_1$ they will produce a local 0 by negating $c_2$, inhibiting the third **V**-gate.

It is simple to see that if both $c_1$ and $c_2$ are 0, then all gates will be inhibited.

This analysis clearly shows that the dual Toffoli gate becomes active, when the *disjunction* –(OR)– of the control signals is *true*. Furthermore, the target line of its Barenco *et al.* model contains only **V**-gates, whereas in the case of classical Toffoli gates the target line contains two **V**-gates and one adjoint **V**-gate [5]}.

*Definition 2*: A basic dual Toffoli gate stays under *mixed polarity*, if it becomes active when both control signals are different.

Fig. 3 shows dual Toffoli gates under mixed polarity, where if a 0 control signal is meant to be effective, then a white triangle will be placed on the control line, in analogy to the "white dots" of the conjunctive case (see e.g. [14]). The "=" sign in Fig. 3 refers to the functionality, not to the structure of the gates. Furthermore, the Barenco *et al.* type of quantum models only illustrate the functionality of the gates. Prevailing "quantum technologies" may not necessarily support negative control of elementary quantum gates.

The gates of Fig. 3 become active when at least one of the controls is effective, i.e. when a white triangle is driven by 0 or a black triangle is driven by 1. The gates become inhibited, when *both* controls are ineffective.

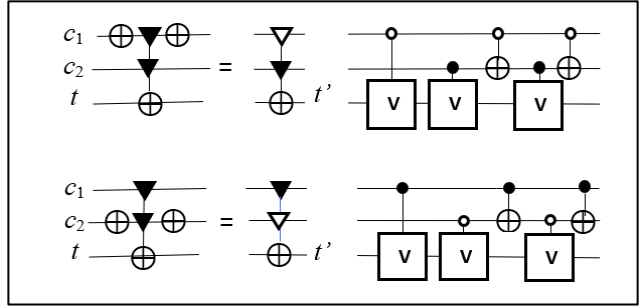

**Fig. 3**: Dual Toffoli gates with mixed polarity and their Barenco *et al.* functional quantum models.

Finally, a dual Toffoli gate, which becomes active when any or both controls have the value 0 is shown in Fig. 4. In this case it is said that the gate has *negative polarity*.

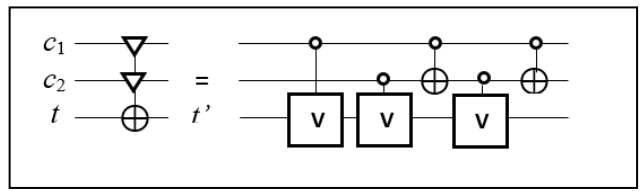

**Fig. 4**: Dual Toffoli gate with negative polarity and its Barenco *et al.* functional quantum model.

As mentioned earlier, the role of the last CNOT gate of the Barenco *et al.* models is the recovering of the initial value of $c_2$, but it does not affect the target output. If this gate is deleted, the target will not be affected, but the output at the middle qubit will change. It is possible to conclude that the modified gate has the behavior of a Peres gate [23], or more precisely, that of a dual Peres gate. From Figs. 3 and 4 it may be seen that to cancel the last CNOT gate, "a similar one" should be added in cascade. (Color consistency should be observed, as in the original Peres gate under mixed polarity). As shown in Fig. 5, the polarity of the gate will affect the output at the middle qubit.

Recall that a Peres gate is not self-inverse. If the inverse is needed, its Barenco *et al.* model is the mirror of that of the original gate.

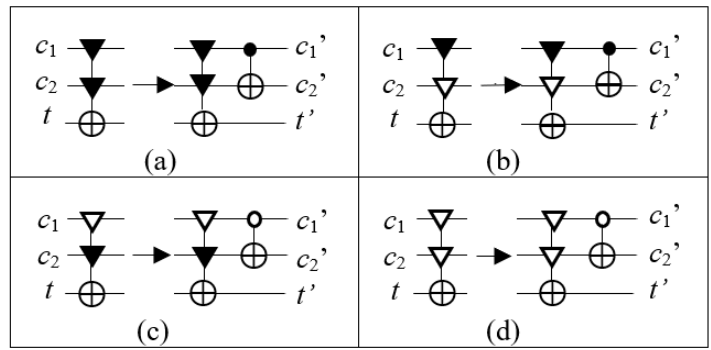

**Fig. 5:** Dual Peres gates under different polarities obtained from dual Toffoli gates
(a), (b): $c'_1 = c_1$ ; $c'_2 = c_1 \oplus c_2$ ; (c), (d): $c'_1 = c_1$ ; $c'_2 = \overline{c}_1 \oplus c_2$
(a) $t' = t \oplus (c_1 \vee c_2)$ ; (b) $t' = t \oplus (c_1 \vee \overline{c}_2)$ ; (c) $t' = t \oplus (\overline{c}_1 \vee c_2)$ ; (d) $t' = t \oplus (\overline{c}_1 \vee \overline{c}_2)$

In early applications using the Peres gate a symbolic representation was done comprising a Toffoli gate and an additional CNOT gate, meant to cancel the last CNOT of the corresponding Barenco *et al.* model (See e.g. [11] Fig. 10). This representation is also used in Fig. 5. Since this representation seemed to be possibly misleading, because the symbolic Peres gate looked more complex than the Toffoli gate instead of being simpler, some authors started to use another symbol for the Peres gate, including a double circle on the bit that should output the sum of the control signals (see e.g. [13]). This last symbol will be adapted for the dual Peres gate, as shown in Fig. 6(c).

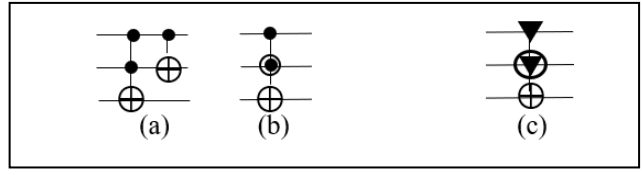

**Fig. 6**: (a), (b) Prevailing symbols for the Peres gate.
(c) Adapted symbol for the dual Peres gate.

With respect to the inverse Peres gate the most frequently used symbol found in the literature consists of a CNOT gate followed by a Toffoli gate, i.e. it is the "mirror" of Fig. 6(a). This is consistent with the Barenco *et al.* quantum model for this gate, since it is the mirror of the model for the original Peres gate. For the case of inverse dual Peres gates, the symbol of Fig. 6(c) will be adapted, by using up-side-down triangles. This is shown in Fig. 7.

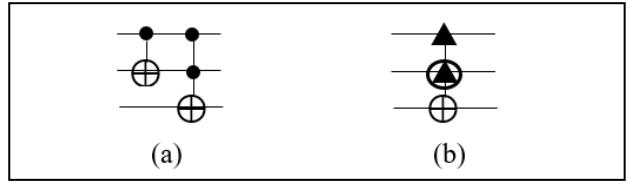

**Fig. 7**: (a) Symbol for the inverse Peres gate.
(b) Symbol for the inverse dual Peres gate.

Still an important structural aspect of dual Toffoli gates has to be considered: the scalability, i.e. the possibility of building multicontrolled dual Toffoli gates, and their decomposability into circuits of basic dual Toffoli / dual Peres gates. A direct realization of a dual Toffoli gate with 3 controls adapted from [20] is shown in Fig. 8, where the gray boxes on the target line represent W-gates. W equals the fourth root of NOT:

$$W = \tfrac{1}{2}\begin{bmatrix} 1 + i^{1/2} & 1 - i^{1/2} \\ 1 - i^{1/2} & 1 + i^{1/2} \end{bmatrix} \qquad [16]$$

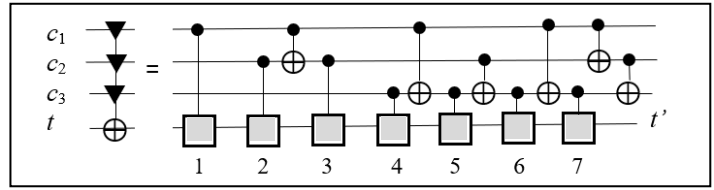

**Fig. 8**: Direct realization of a dual Toffoli gate with 3 controls: $t' = t \oplus (c_1 \vee c_2 \vee c_3)$

**Table 1:** Proof of correctness of the circuit of Fig. 8:

| $c_1$ | 1 | 0 | 1 | 0 | 1 | 0 | 1 |
|---|---|---|---|---|---|---|---|
| $c_2$ | 0 | 1 | 1 | 0 | 0 | 1 | 1 |
| $c_3$ | 0 | 0 | 0 | 1 | 1 | 1 | 1 |
| Active Gates | 1, 3 5, 7 | 2, 3 6, 7 | 1, 2 5, 7 | 4, 5 6, 7 | 1, 3 4, 6 | 2, 3 4, 5 | 1, 2 4, 7 |

From Table 1 it becomes apparent that for any combination of control signals such that $c_1 \vee c_2 \vee c_3 \neq 0$, four W-gates will be active and their cascade –(product)– will generate the expected 3-controlled NOT behavior. If all three control signals have the value 0, then the dual Toffoli gate will be inhibited.

A particular property of this design, which is not based on [5], (but has the same cost –(13)– as the realization in [5]), is the fact that if the value of the controls are expressed in the bit reverse coding, (as shown in Table 1), then where a single 1 appears, the corresponding bit controls its associated W-gate and, where more than one 1 appears, the upper and the lower 1's indicate the positions of controls, such that the upper one indicates the control of a CNOT (placed on the line indicated by the lower 1), that will drive the control of the target gate. As an extension of the simple dual Toffoli gate, the last two CNOT gates recover the control values of $c_1$ and $c_2$. As shown in [21] this method allows a straightforward extension to a larger number of controls. If $k$ controls are considered, the target will have $2^k-1$ equal gates, each one realizing the $(2^{k-1})$-th root of NOT. Since for every control vector –(except the 0 vector)- $2^{k-1}$ target gates become activated, their product returns the expected controlled negation.

Instead of a direct realization of a *k*-controlled dual Toffoli gate, it is possible to realize, as frequently done with classical Toffoli gates, a V-shaped decomposition based on simpler gates following [5], but requiring ancilla lines driven by 0, as shown in Fig. 9.

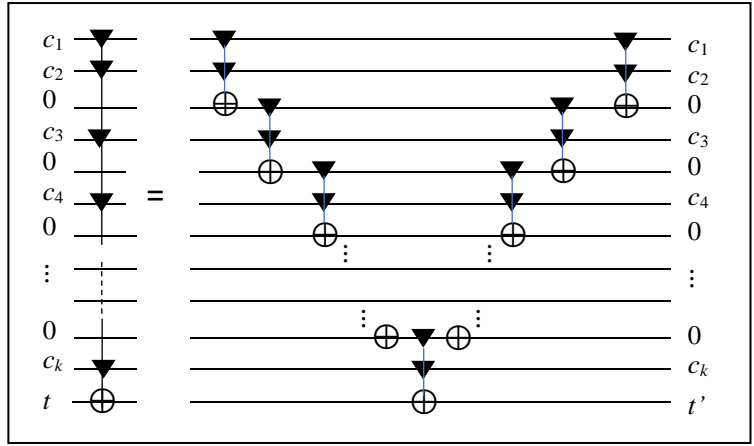

**Fig. 9**: Decomposition of a *k*-controlled dual Toffoli gate based on basic dual Toffoli gates.

The basic dual Toffoli gates at the right hand side of the V-shaped decomposition are meant to recover the control signals. The realization cost may be reduced if the V-components (except for the bottom one) are dual Peres and inverse dual Peres gates, as shown in Fig. 10. (Recall that Peres gates have a quantum cost of 4.)

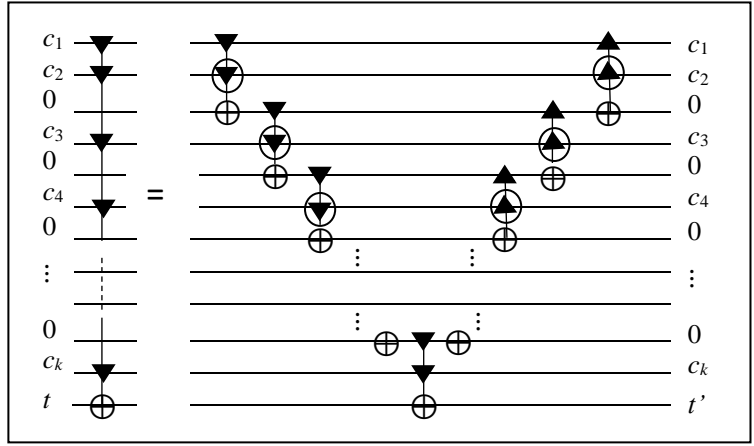

**Fig. 10**: Decomposition of a *k*-controlled dual Toffoli gate based on basic dual Peres gates.

### III. Rewriting rules

Rewriting rules comprise indications on how to move gates within a circuit and replace sub-circuits with simpler ones. (See e.g. [26]). Templates [24] comprise pairs of sub-circuits, where one has the inverse functionality of the other. (Their cascade leads to an identity). The simpler will be used.

In the case of dual Toffoli gates, most rewriting rules may be obtained based on construction, considering that $x \vee y = x \oplus y \oplus xy = x \oplus \bar{x}y = y \oplus x\bar{y}$ and also that $x \vee y = \overline{\bar{x}\bar{y}}$. This is shown in Fig. 11, which, as the basic straight forward rewriting rule, must be read (according to the arrow), from right to left.

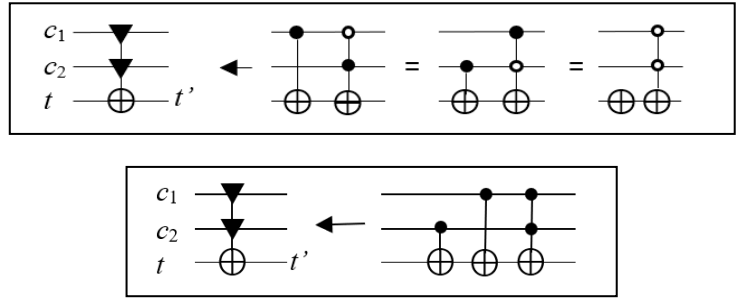

**Fig. 10**: Basic straightforward rewriting rule:

$$\overline{\bar{c}_1 \bar{c}_2} = c_2 \oplus c_1 \bar{c}_2 = c_1 \oplus \bar{c}_1 c_2 = c_1 \vee c_2$$
$$c_1 \oplus c_2 \oplus c_1 c_2 = c_1 \vee c_2$$

**Rule 1**:

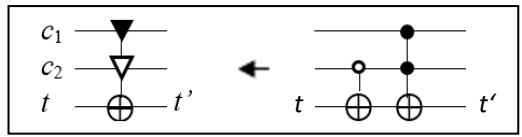

Proof of equivalence: $t' = t \oplus (c_1 \vee \bar{c}_2) = t \oplus c_1 \oplus \bar{c}_2 \oplus c_1\bar{c}_2 = t \oplus \bar{c}_2 \oplus c_1 c_2$

**Rule 2**:

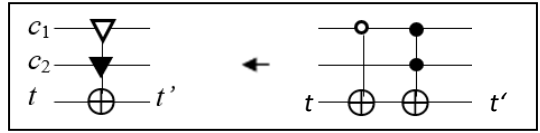

Proof of equivalence: $t' = t \oplus (\bar{c}_1 \vee c_2) = t \oplus \bar{c}_1 \oplus c_2 \oplus \bar{c}_1 c_2 = t \oplus \bar{c}_1 \oplus c_1 c_2$

**Rule 3**:

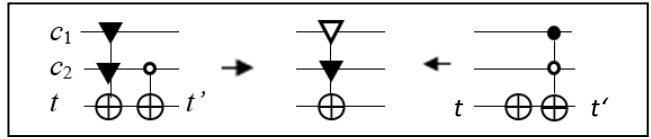

Proof of equivalence: $t' = t \oplus (c_1 \vee c_2) \oplus \bar{c}_2 = t \oplus c_1 \oplus c_2 \oplus c_1 c_2 \oplus \bar{c}_2 = t \oplus 1 \oplus c_1 \bar{c}_2 = t \oplus (\bar{c}_1 \vee c_2)$

**Rule 4**:

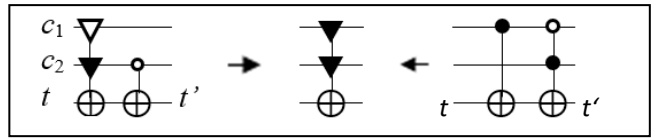

Proof of equivalence: $t' = t \oplus (\bar{c}_1 \vee c_2) \oplus \bar{c}_2 = t \oplus \bar{c}_2 \oplus \bar{c}_1 \oplus c_2 \oplus \bar{c}_1 c_2 = t \oplus 1 \oplus \bar{c}_1 \oplus \bar{c}_1 c_2 = t \oplus c_1 \oplus \bar{c}_1 c_2$

But also $\quad t \oplus 1 \oplus \bar{c}_1 \oplus \bar{c}_1 c_2 = t \oplus 1 \oplus \bar{c}_1 \bar{c}_2 = t \oplus (c_1 \vee c_2)$

**Rule 5**:

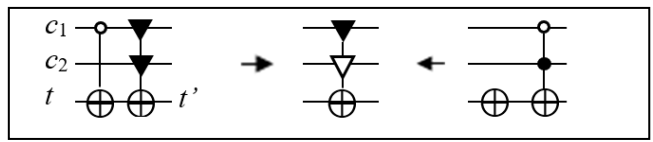

Proof of equivalence: $t' = t \oplus \bar{c}_1 \oplus (c_1 \vee c_2) = t \oplus \bar{c}_1 \oplus c_1 \oplus c_2 \oplus c_1 c_2 = t \oplus 1 \oplus c_2 \oplus c_1 c_2 = t \oplus 1 \oplus \bar{c}_1 c_2$

But also $\quad t \oplus 1 \oplus c_2 \oplus c_1 c_2 = t \oplus (c_1 \vee \bar{c}_2)$

**Rule 6:**

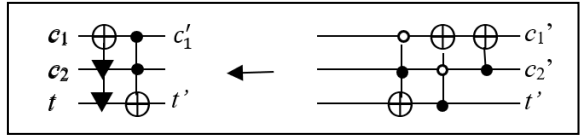

Proof of equivalence: $c_1' = c_1 \oplus (c_2 \vee t) = c_1 \oplus c_2 \oplus t \oplus c_2 t = c_1 \oplus c_2 \oplus \bar{c}_2 t$ ; $c_2' = c_2$

$$t' = t \oplus c_1' c_2 = t \oplus (c_1 \oplus c_2 \oplus \bar{c}_2 t) c_2 = t \oplus c_1 c_2 \oplus c_2 = t \oplus \bar{c}_1 c_2$$

However, if both sides of $t' = t \oplus \bar{c}_1 c_2$ are multiplied by $\bar{c}_2$, then

$$\bar{c}_2 t' = \bar{c}_2 t$$

and $\qquad c_1' = c_1 \oplus c_2 \oplus \bar{c}_2 t'$

The circuit based on a dual Toffoli gate and a classical Toffoli gate gives the simplest realization.

**Rule 7**:

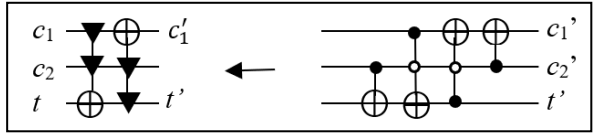

Proof of equivalence: $t' = t \oplus (c_1 \vee c_2) = t \oplus c_1 \oplus c_2 \oplus c_1 c_2 = t \oplus c_2 \oplus c_1 \bar{c}_2$

$c_1' = c_1 \oplus (t' \vee c_2) = c_1 \oplus t' \oplus c_2 \oplus t' c_2 = c_1 \oplus c_2 \oplus t' \bar{c}_2$

**Rule 8**:

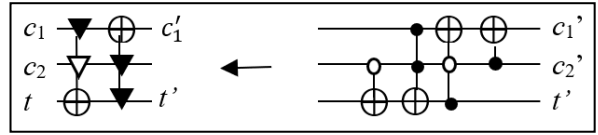

Proof of equivalence: $t' = t \oplus (c_1 \vee \bar{c}_2) = t \oplus c_1 \oplus \bar{c}_2 \oplus c_1 \bar{c}_2 = t \oplus \bar{c}_2 \oplus c_1 c_2$

$c_1' = c_1 \oplus (t' \vee c_2) = c_1 \oplus t' \oplus c_2 \oplus t' c_2 = c_1 \oplus c_2 \oplus t' \bar{c}_2$

**Rule 9**:

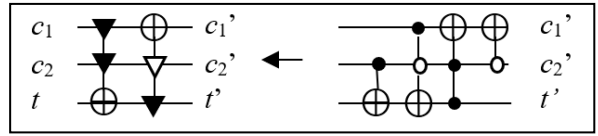

Proof of equivalence: $t' = t \oplus (c_1 \vee c_2) = t \oplus c_1 \oplus c_2 \oplus c_1 c_2 = t \oplus c_2 \oplus c_1 \bar{c}_2$

$c_1' = c_1 \oplus (t' \vee \bar{c}_2) = c_1 \oplus t' \oplus \bar{c}_2 \oplus t' \bar{c}_2 = c_1 \oplus \bar{c}_2 \oplus t' c_2$

**Rule 10**:

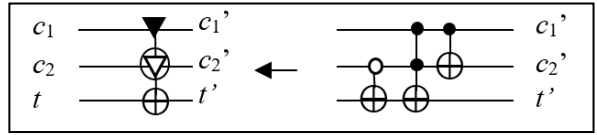

Proof of equivalence: $c_1' = c_1$ ; $c_2' = c_1 \oplus c_2$

$$t' = t \oplus (c_1 \vee \bar{c}_2) = t \oplus c_1 \oplus \bar{c}_2 \oplus c_1 \bar{c}_2 = t \oplus \bar{c}_2 \oplus c_1 c_2$$

**Rule 11**:

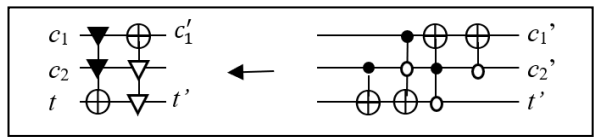

Proof of equivalence: $t' = t \oplus (c_1 \vee c_2) = t \oplus c_1 \oplus c_2 \oplus c_1 c_2 = = t \oplus c_2 \oplus c_1 \bar{c}_2$ ; $c'_2 = c_2$

$c'_1 = c_1 \oplus (\overline{t'} \vee \bar{c}_2) = c_1 \oplus \overline{t'} \oplus \bar{c}_2 \oplus \overline{t'}\bar{c}_2 = c_1 \oplus \bar{c}_2 \oplus \overline{t'} c_2$

**Rule 12**:

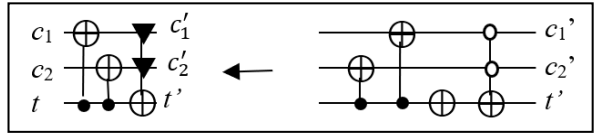

Proof of equivalence: $c'_1 = t \oplus c_1$ ; $c'_2 = t \oplus c_2$

$t' = t \oplus (c'_1 \vee c'_2) = t \oplus ((t \oplus c_1) \vee (t \oplus c_2)) =$

$= t \oplus t \oplus c_1 \oplus t \oplus c_2 \oplus (t \oplus c_1)(t \oplus c_2) =$

$= t \oplus c_1 \oplus c_2 \oplus t \oplus tc_1 \oplus tc_2 \oplus c_1 c_2 = c_1 \bar{t} \oplus c_2 \bar{t} \oplus c_1 c_2$

In the circuit at the right:

$t' = \bar{t} \oplus \overline{c_1'}\,\overline{c_2'} = \bar{t} \oplus (\bar{t} \oplus c_1)(\bar{t} \oplus c_2) = c_1 \bar{t} \oplus c_2 \bar{t} \oplus c_1 c_2$

**Rule 13**:

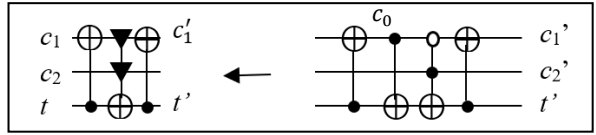

Proof of equivalence: Let $c_0 = t \oplus c_1$

$t' = t \oplus (c_0 \vee c_2) = t \oplus c_0 \oplus c_2 \oplus c_0 c_2 = t \oplus c_0 \oplus \bar{c}_0 c_2$

$c_1' = c_0 \oplus t'$

The first three gates of the circuit at the right realize *t'* and the last CNOT realizes $c_1' = c_0 \oplus t'$.

**Rule 14**:

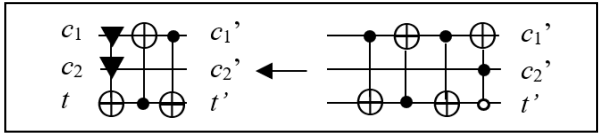

Proof of equivalence: Let $t_0 = t \oplus (c_1 \vee c_2) = t \oplus c_1 \oplus c_2 \oplus c_1 c_2$

$c'_1 = t_0 \oplus c_1 = t \oplus c_1 \oplus c_2 \oplus c_1 c_2 \oplus c_1 = t \oplus \bar{c}_1 c_2$

$t' = t_0 \oplus c'_1 = c_1$

Notice that in the circuit at the right the first three CNOT gates swap $c_1$ and *t*.

These 14 rewriting rules do not cover all possible simplifications, but should be considered as motivating examples offering new possibilities for post-processing of prevailing Toffoli circuits. This may be considered as a "bottom-up" approach. A direct approach has been integrated in an evolutionary design system [6] by including the dual gates in the gates library, obtaining positive results. A "top-down" approach was suggested in [18] starting from the specification of a function as a polynomial and searching for sub-expressions of the type "x $\oplus$ y $\oplus$ xy" or equivalents. Polynomial expressions may be obtained from the value vector of a function with the Reed-Muller transform (see e.g. [10]).

The next step will be the study of dual gates at the level of their Clifford-T model and evaluate their performance in the context of the IBM-QX quantum computers [1], [3]. Fig. 11 shows a possible Clifford-T realization of a dual Toffoli gate based on the equality $x \vee y = x \oplus \bar{x}y$, (recall Fig. 10), and an $\bar{x}y$ gate with optimal T-count of 7 and T-depth of 4, [2]. Fig 12 shows a possible Clifford-T realization of a dual Peres gate based on a $1 \oplus \bar{x}\bar{y}$ gate from [2] followed by a CNOT gate.

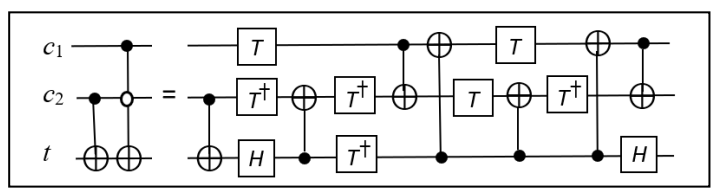

**Fig. 11**: Clifford-T realization of a dual Toffoli gate adapted from [2]

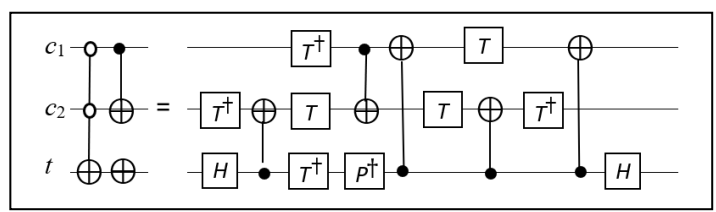

**Fig. 12**: Clifford-T realization of a dual Peres gate adapted from [2]

## IV. Closing Remarks

Dual Toffoli and Peres gates have been presented and their functionality introduced, based on an adapted Barenco *et al.* model for the classical Toffoli gate. A possible way of using these gates to improve existing reversible/quantum Toffoli circuits in form of rewriting rules, are discussed. Prevailing circuits may be improved with the rewriting rules by reducing the number of gates and quantum cost. Some examples on the realization of benchmarks may be found in [6]. Finally, a possible Clifford-T realization of dual Toffoli and Peres gates with minimum T-count is shown, which may be considered when optimized circuits are mapped to an IBM-QX quantum computer.

## References

[1] Abdessaied, N., Amy M., Soeken M., Drechsler R., "Technology mapping of reversible circuits to Clifford+T quantum circuits." *Proc. 2016 46th Int. Symp. on Multiple-Valued Logic (ISMVL)*, 150–155, IEEE Press, 2016.
[2] Amy M., Maslov D., Mosca M., Roetteler M., "A meet-in-the-middle algorithm for fast synthesis of depth-optimal quantum circuits," *IEEE Trans. Computer-Aided Design of Integrated Circuits and* Systems, 32, 818-830, 2013.
[3] Almeida A.A.A., Dueck G.W., Rodrigues da Silva A.C., "Efficient Realization of Toffoli and NCV Circuits for IBM QX Architectures," M.K. Thomsen and M. Soeken (Eds.), RC 2019, LNCS 11497, 131-145, Springer, 2019.
[4] Cheng C.S., Singh A.K., "Heuristic Synthesis of Reversible Logic – A Comparative Study," *Theoretical and Applied Electrical Engineering*, 12, (3), 210-225, 2014.
[5] Barenco A., Bennett C. H., Cleve R., Di Vincenzo D. P., Margolus N., Shor P., Sleator T., Smolin J. A. L. L., Weinfurter H., "Elementary gates for quantum computation," *Phys. Rev.* A 52, 3457–3467, 1995.
[6] Hadjam F.Z., Moraga C.: "RIMEP2. Evolutionary Design of Reversible Digital Circuits," *ACM Jr. of Emerging Technologies in Computing Systems*, 11(3), 27:1-27:23, DOI: http://dx.doi.org/10.1145/2629534, 2014.
[7] Hadjam F.Z., Moraga C.: "A symbolic calculus for a class of quantum computing circuits." *Electronics Letters*, 51, (9): 682-683, 2015.
[8] Hadjam F., Moraga C.: "Distributed RIMEP2: A Comparative Study between a Hierarchical Model and the Islands Model in the context of reversible circuits design." In *Proc. Int. Workshop on Boolean Problems*, 13-20. Press T.U. Freiberg, 2016.
[9] Hung W.N.N., Song X., Yang G., Yang J., Perkowski M., "Quantum Logic Synthesis by Symbolic Reachability Analysis," in *Proc. DAC 2004*, 838-841, ACM Press, 2004.
[10] Karpovsky M.G., Stankovic R.S., Astola J.T, *Spectral Logic and its Applications for the Design of Digital Devices.* John Wiley, 2008.
[11] Lukas M. *et al.*, "Evolutionary Approach to Quantum and Reversible Circuits Synthesis," *Artificial Intelligence Review* 20, 361-417, 2003.
[12] Maslov D., Dueck G. W., Miller D. M., "Fredkin/Toffoli Templates for Reversible Logic Synthesis," in *Proc. ICCAD-2003*, ACM Press, 2003.
[13] Maslov D., Dueck G. W., Miller D. M., "Toffoli network synthesis with templates." *IEEE Trans. CAD Integrated Circuits and Systems* 24 (6), 807–817, 2005.
[14] Maslov D., Dueck G. W., Miller D. M., Negrevergne C., "Quantum Circuit Simplification and Level Compaction," *IEEE Trans. CAD of Integrated Circuits and Systems*, 27 (3), 436-444, 2008.
[15] Miller D.M., Dueck G.W., "Spectral Techniques for Reversible Logic Synthesis," in *Proc. Midwest Symp. on Circuits and Systems 2002.* IEEE Press, 2002.
[16] Miller D.M.. Private communication, 2011.

[17] Mohammadi M., "Efficient Genetic Based Methods for Optimizing the Reversible and Quantum Logic Circuits," *J. Advances in Computer Research Quarterly,* 3, (3), 85-96, 2012.

[18] Moraga C., "Hybrid GF(2)-Boolean Expressions for Quantum Computing Circuits," A. De Vos and R. Wille (Eds.), RC 2011, LNCS 7165, 54-63, Springer, 2012.

[19] Moraga C., "Toffoli gates with Multiple Mixed Control Signals and no Ancillary Lines." B. Steinbach (Ed.) *Recent Progress in the Boolean Domain*, 359-368. ISBN 978-1-4438-5638-6, Cambridge Scholars Publishing, Newcastle upon Tyne, U.K., 2014

[20] Moraga C., "Using negated control signals in quantum computing circuits," *Facta Universitatis*, Series Electronics and Energetics, 24, (3), 423-435. Press University of Niš, Serbia, 2011.

[21] Moraga C., "Mixed polarity reversible Peres gates," *Electronics Letters* 50, (14), 987-989, 2014.

[22] Muthukrishnan A., Stroud C. "Multivalued logic gates for quantum computation." *Phys. Rev. A*, 62, 52309, (2000)

[23] Peres A., "Reversible Logic and Quantum Computers." *Physic Review*, 32, (6), 3266–3276, 1985.

[24] Rahman M., Dueck G.W.. "An algorithm to find quantum Templates", in *Proc. IEEE Congress on Evol. Comp.*, 623-629, IEEE Press, 2012.

[25] Saeedi M., Markov I.L., "Synthesis and optimization of reversible circuits – a survey." *Comput. Surveys,* arXiv 1110.2574v1.

[26] Soeken M., Thomsen M.K., "White Dots *do* Matter: Rewriting Reversible Logic Circuits," G.W. Dueck and D.M. Miller (Eds.), RC 2013, LNCS 7948, 196-208, Springer, 2013.

[27] Soeken M., Dueck G.W., Miller M.D., "A fast symbolic transformation-based algorithm for reversible logic synthesis," Devitt S., Lanese I., (Eds.). *Reversible Computation LNCS* 9720, 307-321, Springer, 2016.

[28] Stojković S., Stanković M.M., Moraga C.: Complexity reduction of Toffoli networks based on FDD. *Facta Universitatis. Series Electronics and Energetics*, 28 (2), 251-262, 2015.

[29] Szyprowski M., Kerntopf P., "Reducing quantum cost in reversible Toffoli circuits." In *Proc. Reed-Muller 2011 Workshop*. 127–136. arXiv:1105.5831.

[30] Toffoli T., Reversible computing. Tech. memo MIT/LCS/TM-151, MIT Lab for Comp. Sci., 1980.

[31] Toffoli T., Reversible Computing, J.W. Baker and J. van Leeuwen (Eds.), ALP 1980, LNCS 84, 6732-644, Springer, 1980.

[32] Wegner I., *The Complexity of Boolean Functions.* John Wiley and Sons, New York, 1987.